\documentclass[twocolumn,superscriptaddress,pre]{revtex4}
\usepackage{epsfig}
\usepackage{amssymb,amsfonts,amsmath,color}

\include{epsf}

\def\(({\left(}
\def\)){\right)}                       
\def\[[{\left[}
\def\]]{\right]}

\newcommand{\ben}{\begin{enumerate}}
\newcommand{\een}{\end{enumerate}}
\newcommand{\bit}{\begin{itemize}}
\newcommand{\eit}{\end{itemize}}

\newcommand{\<}{\langle}
\renewcommand{\>}{\rangle}

\newcommand{\beq}{\begin{equation}}
\newcommand{\eeq}{\end{equation}}
\newcommand{\bea}{\begin{eqnarray}}
\newcommand{\eea}{\end{eqnarray}}

\renewcommand{\phi}{\varphi}
\newcommand{\SI}{}

\usepackage{amsmath}
\usepackage{amssymb}
\usepackage{amsthm}
\usepackage{amsfonts}

\begin{document}
\title{A small-correlation expansion to quantify information in noisy sensory systems}
\author{Gabriel Mahuas}
\affiliation{Institut de la Vision, Sorbonne Université, CNRS, INSERM, 17 rue Moreau, 75012, Paris, France}
\affiliation{Laboratoire de physique de \'Ecole normale sup\'erieure,
  CNRS, PSL University, Sorbonne University, Universit\'e Paris-Cit\'e, 24 rue Lhomond,
  75005 Paris, France}
\author{Olivier Marre}
\affiliation{Institut de la Vision, Sorbonne Université, CNRS, INSERM, 17 rue Moreau, 75012, Paris, France}
\author{Thierry Mora}
\thanks{To whom correspondence should be sent. These authors
  contributed equally.}
\affiliation{Laboratoire de physique de \'Ecole normale sup\'erieure,
  CNRS, PSL University, Sorbonne University, Universit\'e Paris-Cit\'e, 24 rue Lhomond,
  75005 Paris, France}
\author{Ulisse Ferrari}
\thanks{To whom correspondence should be sent. These authors
  contributed equally.}
\affiliation{Institut de la Vision, Sorbonne Université, CNRS, INSERM, 17 rue Moreau, 75012, Paris, France}

\linespread{1}

\begin{abstract}
Neural networks encode information through their collective spiking activity in response to external stimuli. This population response is noisy and strongly correlated, with complex interplay between correlations induced by the stimulus, and correlations caused by shared noise. Understanding how these correlations affect information transmission has so far been limited to pairs or small groups of neurons, because the curse of dimensionality impedes the evaluation of mutual information in larger populations.
Here we develop a small-correlation expansion to compute the stimulus information carried by a large population of neurons, yielding interpretable analytical expressions in terms of the neurons' firing rates and pairwise correlations. We validate the approximation on synthetic data and demonstrate its applicability to electrophysiological recordings in the vertebrate retina, allowing us to quantify the effects of noise correlations between neurons and of memory in single neurons.

\end{abstract}

\maketitle

Networks of neurons from sensory systems are characterized by strong correlations that shape their collective response to stimuli \cite{Brivanlou98,Trenholm14,Schneidman06,Shlens08,Trong08,Ala11,Deweese04,Lin15,Kohn05,Montani07,Smith08,PonceAlvarez13,Franke16,Zylberberg16}. These correlations have two sources \cite{Brivanlou98}: {\em stimulus correlations}, which originate from shared or correlated stimuli that affect the mean activities of different neurons in a concerted way; and {\em noise correlations}, which stem from network interactions that couple noise across cells. These two sources of correlations impact how well the population encodes stimulus information, and detailed investigations have explored this effect both experimentally  \cite{Montani07,Smith08,PonceAlvarez13,Franke16,Zylberberg16,Meister95,Nirenberg01,Pillow08,Ruda20,Hazon22,Boffi22} and theoretically \cite{Zohary94,Brunel98,Abbott99,Panzeri99,Sompolinsky01,Pola03,Tkacik10,Ecker10,Ecker11,Hu14,Moreno-bote14}, showing a wide variety of scenarios in which noise correlations could either hurt or improve information transmission (see \cite{Azeredo21} for a recent review).

While geometric arguments about the structure of stimulus and noise correlations can help interpret and evaluate the impact of their interplay on information transmission for pairs or small groups of cells \cite{Hu14,Azeredo21}, specific challenges arise when dealing with large populations of cells. A common way to quantify these effects is by computing the mutual information between the stimulus and the activity of the whole population.
However, attempts at quantifying this information are inherently limited by the curse of dimensionality, whereby the size of the state space to be sampled grows exponentially with the system's size. Models based on the principle of maximum entropy have been proposed to build explicit probabilistic models of the collective activity of many neurons, based on mean spike rates and correlation functions \cite{Jaynes82,Schneidman06,Tang08,Roudi09,Shimazaki12,Watanabe13,Granot-Atedgi13,Tkacik13b,Tkacik14,Tkacik15,Cofre14,Mora15,Shimazaki15,Tavoni17,Meshulam17,Ferrari17a,Nghiem18,Cayco18,Delamare22}. These distributions map onto known models of statistical mechanics, and can be used to evaluate entropies as well as mutual informations.

In this paper, we leverage these techniques from statistical physics to compute a small correlation expansion of the mutual information \cite{Sessak09}.
This approach outperforms previous approximations of the mutual information and is computationally efficient. The resulting formulas, while implicitly relying on the maximum entropy assumption, are expressed in a model-free manner as a function of the experimental observables, yielding a very simple picture of how correlations affect information encoding in sensory systems.
We test this approach on synthetic data and showcase it on real electrophysiological recordings from the retina, to illustrate how it can be used to quantify the effect of noise correlations between neurons in the population as well as the effect of temporal correlations in single neurons.

\begin{figure}[th!]
\centering
\includegraphics[width=\columnwidth]{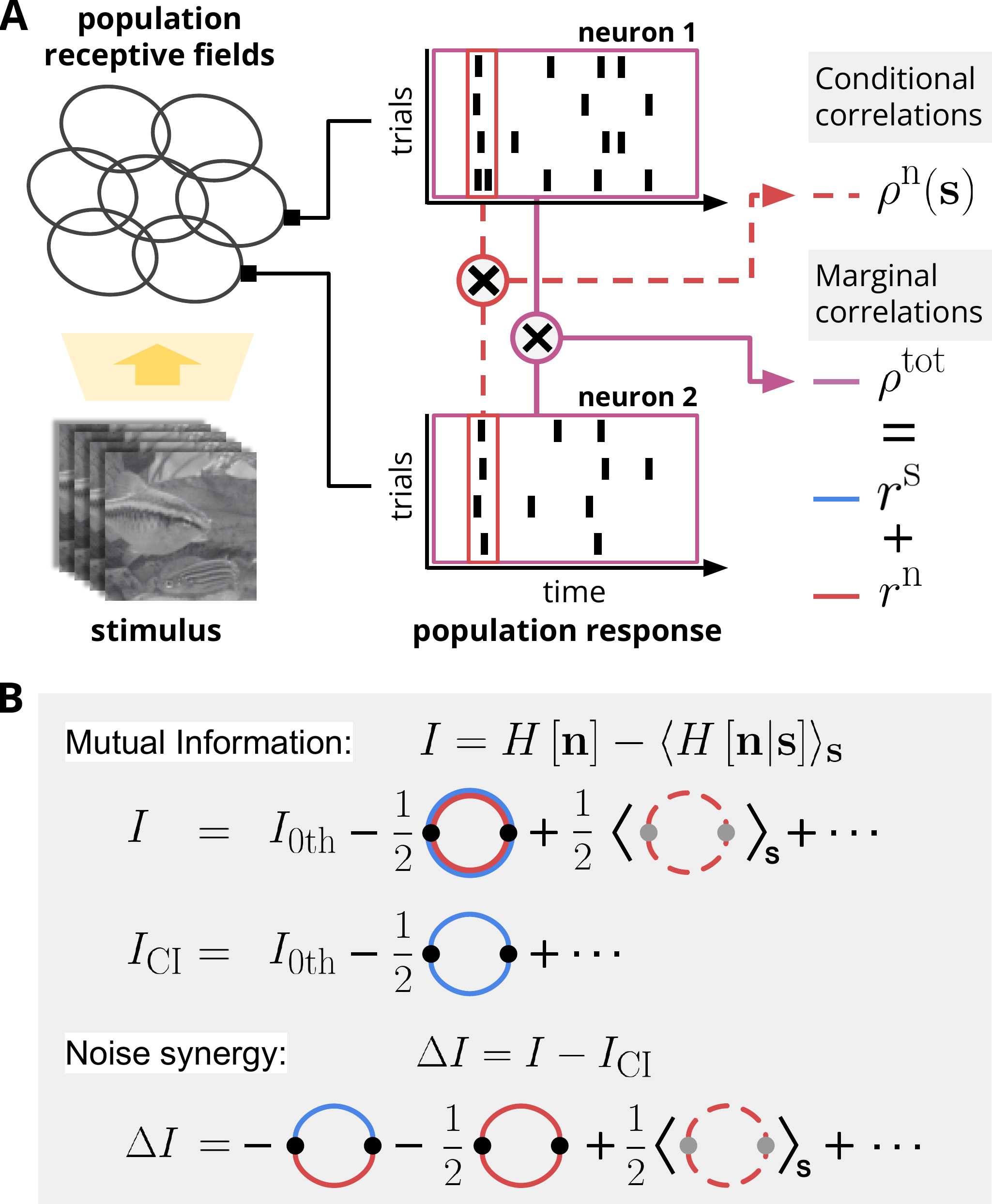}
\caption{\textbf{How correlations affect information.}
A) Visual stimuli drive the noisy response of sensory neurons (spikes, represented by vertical ticks). 
We call $\rho_{ij}^{\rm n}({\bf s})$ the pairwise Pearson correlation of binary activities $n_i,n_j$ between cells $i$ and $j$ in a short window $\Delta t$, conditioned on past stimulus $\bf s$. The unconditioned Pearson correlation, $\rho^{\rm tot}$,
can be decomposed into stimulus and noise contributions, $r^\text{s}$ and $r^\text{n}$.
B) We estimate the mutual information $I$ between stimulus $\bf s$ and response $\bf n$ through a small correlation expansion \cite{Sessak09}. 
Noise synergy $\Delta I$ estimates the impact of noise correlations onto mutual information, and is defined as the gain in information relative to the conditionally independent information $I_\text{CI}$, corresponding to no noise correlations (obtained by shuffling neural responses across repetitions).
In the diagrammatic expansions, each line corresponds to a correlation term; double lines are sums of two correlations; multiple lines connecting the same two points are multiplied.
}
\label{f:1}
\end{figure} 

{\em Small correlation expansion of the mutual information.}
The collective response of a neural network of size $N$ can be described by the neuronal activities ${\bf n}=(n_1,\ldots,n_N)$, taking value 0 or 1 depending on whether the neuron spikes or not within a short time window $\Delta t $ (typically 10\,--\,20 ms). 
In general, because of processing delays and adaptation, the response is a stochastic function $P(\bf n|\bf s)$ of the history of the stimulus $\bf s$ up to the response.
The mutual information $I({\bf n},{\bf s})$ quantifies the amount of information conveyed by the neural response about the stimulus \cite{Shannon48,Rieke99}. Since it is expressed as a difference of entropies
$I=H[\mathbf{n}] - \<H[{\bf n}|{\bf s}]\>_{\bf s}$, where $H[x]=-\sum_x P(x)\ln P(x)$,
its quantification requires good entropy estimators. Direct estimation methods from data exist, and can be applied for relatively small groups of neurons \cite{Strong98}. However, the estimation problem quickly becomes intractable as the number of neurons increases and the size of the response space grows exponentially. To deal with large networks, we thus developed a method based on a small correlation expansion of entropies \cite{Sessak09}, which allows us to express them as analytical functions of the empirical correlations.

We start by assuming that both $P({\bf n})$ and $P({\bf n}|{\bf s})$ follow the form of maximum entropy models consistent with empirical pairwise covariances and spike rates.
Later we will discuss the limitations of this assumption.
The total covariance between two cells $i$ and $j$ across stimuli, $C_{ij}\equiv\mathrm{Cov}(n_i,n_j)$, can be decomposed into two contributions corresponding to the effects of stimulus and noise: $C_{ij}=C_{ij}^{\rm s}+\<C_{ij}^{\rm n}({\bf s})\>_{\bf s}$, with $C_{ij}^{\rm s}\equiv\mathrm{Cov}_{\bf s}(\<n_i\>_{n_i |\bf s}, \<n_j\>_{n_j |{\bf s}})$, $C_{ij}^{\rm n}({\bf s})\equiv\mathrm{Cov}(n_i,n_j|{\bf s})$, which can be computed from the response to repeated presentations of the same stimulus.
Likewise, the Pearson correlation coefficient $\rho_{ij}^\text{tot}\equiv C_{ij}/\sqrt{C_{ii}C_{jj}}$ can also be decomposed into stimulus- and noise-induced contributions: $\rho_{ij}^\text{tot} = r_{ij}^\text{s} + r_{ij}^\text{n}$ (Fig.~\ref{f:1}A), with $r_{ij}^\text{s}\equiv C^{\rm s}_{ij}/\sqrt{C_{ii}C_{jj}}$ and $r_{ij}^\text{n}\equiv C^{\rm n}_{ij}/\sqrt{C_{ii}C_{jj}}$. Note however that these two terms are not proper correlation coefficients because of the normalization. Stimulus correlations may instead be quantified by $\rho_{ij}^\text{s}\equiv C^{\rm s}_{ij}/\sqrt{C^{\rm s}_{ii}C^{\rm s}_{jj}}$, and noise correlations in a stimulus-dependent manner through: $\rho_{ij}^\text{n}({\bf s})\equiv C^{\rm n}_{ij}({\bf s})/\sqrt{C^{\rm n}_{ii}({\bf s}) C^{\rm n}_{jj}({\bf s})}$.

Following the approach of Sessak and Monasson \cite{Sessak09}, we can expand the entropy of the maximum entropy models---and thus the mutual information---at small values of the correlation parameter, $I=I_{\rm 0th}+I_{\rm 1st}+I_{\rm 2nd}+\ldots$ (App.~A\SI).
The leading order of this expansion is the sum of the information carried by each neuron: $I_{\rm 0th}= \sum_i\left[ H[{n_i}] -\<H[n_i|{\bf s}]\>_{\bf s}\right]$.
The first order term vanishes, while the second one reads (Fig.~\ref{f:1}B, App. A\SI): 
\begin{equation}\label{eq:MI_dep_2cd}
I_{\rm 2nd} = - \frac{1}{2} \sum_{i<j} \left( {\rho_{ij}^{\rm tot}}^2 - \<\rho^{\rm n}_{ij}(\mathbf{s})^2 \>_{\bf s} \right).
\end{equation}
We can compute higher-order terms using Feynman diagrams rules \cite{Kuhn22}, but they quickly become unwieldy. 
However, some of these terms can be re-summed to yield a better approximation of the mutual information than \eqref{eq:MI_dep_2cd} in terms of first and second order moments~\cite{Sessak09} (App. B\SI):
\begin{align}\label{eq:MI_dep_resumation}
I &\approx I_{\rm 0th} + I_{\rm  pairs}  + I_{\mathcal{G}} - I_{\rm dbl}.
\end{align}
$I_\text{\rm  pairs}$ is the sum of the mutual information gains (with respect to single cells) of each pair $(i,j)$ calculated one by one, ignoring the rest of the network. 
$I_{\mathcal{G}}$ is the mutual information gain computed through a mean-field (or \textit{loop}) approximation \cite{Cocco11,Sessak09}, which is equivalent to assuming that all fluctuations (stimulus and noise) are Gaussian:
\begin{align}\label{eq:MI_loop}
\begin{split}
I_{\mathcal{G}} &= \frac{1}{2}\log\left( |\rho^{\rm tot}| \right) - \frac{1}{2}\<\ln\left( |\rho^{\rm n}(\mathbf{s})| \right)\>_{\bf s},
\end{split}
\end{align}
where $|\rho|$ denotes the determinant of the correlation matrix.
Finally $I_\text{dbl}$ corrects for terms that are double-counted in $I_{\rm  pairs}$ and $I_{\mathcal{G}}$.

{\em Noise synergy.}
These expansions can be used to investigate the impact of noise correlations on information transmission.
We define the \textit{noise synergy}, $\Delta I \equiv I - I_\text{CI}$, as the gain in information relative to the conditionnally independent case (Fig.~\ref{f:1}B, bottom line). $I_{\rm CI}$ can be computed in practice by shuffling the response of individual neurons across repetitions of the same stimulus, which preserves stimulus correlations but destroys noise correlations. At second order we obtain (App. A\SI):
\begin{align}\label{eq:noiseSyn}
\Delta I \approx \sum_{i<j} \left[ -r_{i,j}^{\mathrm{n}}r_{i,j}^{\mathrm{s}} + \frac{1}{2}\left(\left<{\rho_{i,j}^{\mathrm{n}}(\mathbf{s})}^2\right>_{\mathbf{s}} - {r_{i,j}^{\mathrm{n}}}^2\right) \right].
\end{align}
This expression shows how noise synergy depends on noise correlations through $r^{\mathrm{n}}$ and $\mathbf{\rho}^{\mathrm{n}}$. 
The first term is positive when noise and stimulus correlations have opposite signs.
This effect is known in the literature as the \textit{sign-rule} \cite{Hu14} and can be interpreted in terms of the whitening of the output power spectrum: it is beneficial for the network to ``cancel out'' input correlations by adding noise correlations of opposite sign, in order to approach a uniformly distributed output, thereby increasing output entropy and information.
The second term of \eqref{eq:noiseSyn}, which is of second order in the noise correlation parameter, can be either positive or negative in general.
However, in the particular case of noise correlations independent of the stimulus, $\rho^\text{n}({\bf s})=\rho^{\rm n}$, the Cauchy-Schwarz inequality guarantees its non-negativity (see App.~A\SI{} for a proof). This implies that noise correlations may be beneficial even when the sign rule is violated.
Noise synergy can also be computed using the re-summed entropies of \eqref{eq:MI_dep_resumation}. The formulas are slightly more involved and are reported in App.~B\SI.

\begin{figure}[t!]
\centering
\includegraphics[width=\columnwidth]{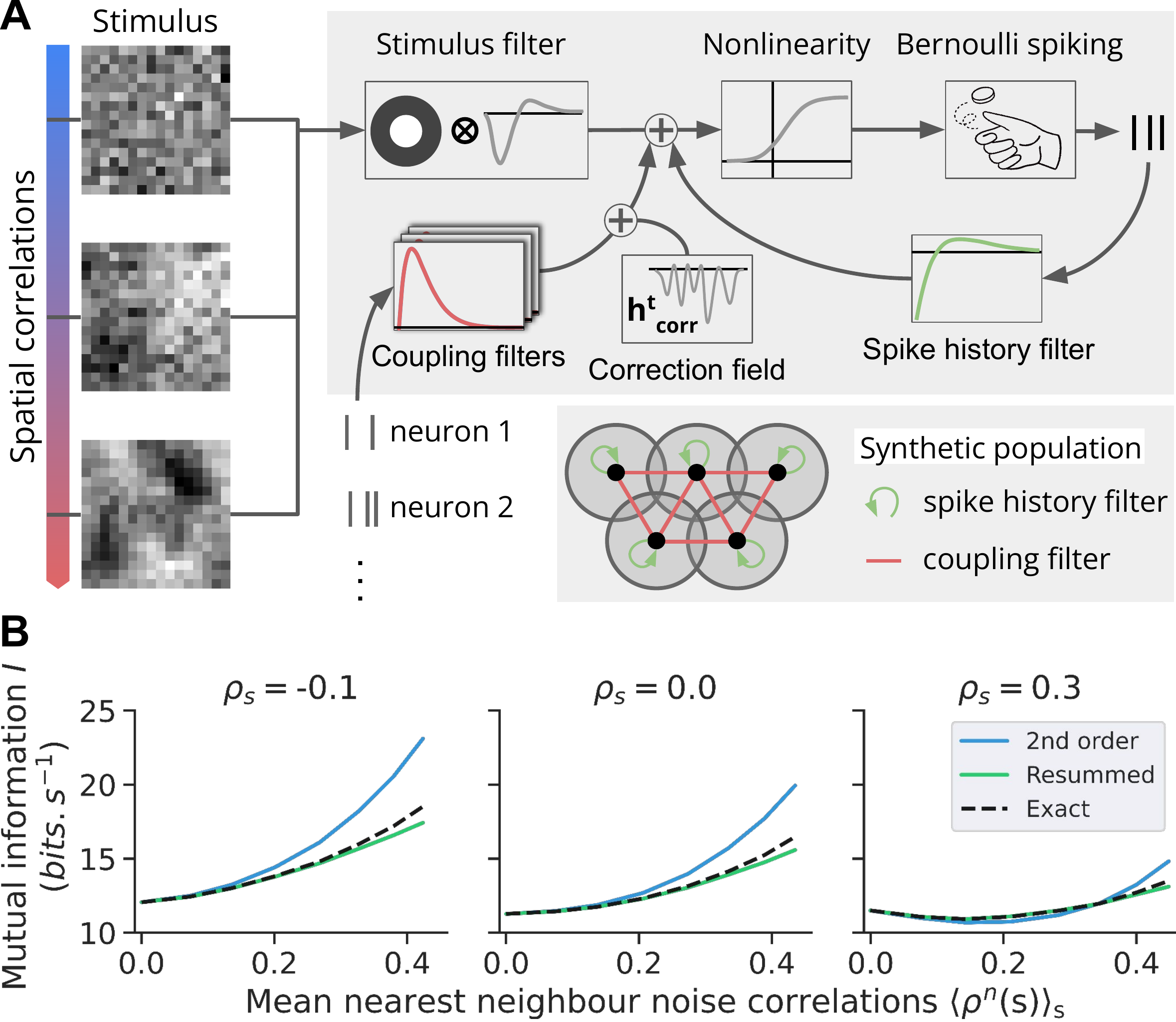}
\caption{
  A) A spatially correlated random stimulus activates a network of 5 neurons according to a generalized linear model defined by stimulus, coupling, and spike-history filters.
B) Exact, second order (\ref{eq:MI_dep_2cd}) and re-summed (\ref{eq:MI_dep_resumation}) values of the mutual information for various strengths of the noise and stimulus correlations (averaged over all pairs of neighbors).
}
\label{f:2}
\end{figure}

{\em Numerical test on synthetic data.}
To test our approximations \eqref{eq:MI_dep_2cd} and \eqref{eq:MI_dep_resumation}, we built a generalized linear model to mimic the response of a small population of five retinal neurons with nearest-neighbor interactions (Fig.~\ref{f:2}A) for which mutual informations could be estimated exactly.
The stimulus is modeled as a random Gaussian field sampled at 100Hz, with varying spatial correlations, allowing us to tune the strength of stimulus correlations (App.~C\SI).
The stimulus is convolved with a linear filter consisting of a Gaussian receptive field with biphasic temporal kernel \cite{Mahuas20} (App.~C\SI). The mean spike rate is controled by the result of this convolution, to which the effect of its own spiking history is added, through a non-linear function. In addition, the past activities of its neighbors control the stochastic part of firing, through coupling filters (the mean effect of which is subtracted from the average rate, see App.~C\SI).
This strategy allows us to tune noise correlations while keeping stimulus correlation constant. Importantly, this model is mathematically inconsistent with the maximum entropy assumption. It thus allows us to test for both the appropriateness of the maximum entropy approximation in the context of state-of-the-art spiking models, and the accuracy of the small-correlation expansion.

We then computed the exact mutual information between stimulus and response using exhaustive numerical simulations, and compared it with the predictions of our approximations (Fig.~\ref{f:2}B). 
We observed an excellent agreement between numerical calculations and analytical expressions, in particular for
the re-summed mutual information \eqref{eq:MI_dep_resumation}. Although less accurate, the second order approximation \eqref{eq:MI_dep_2cd} still provided fair estimates for a wide range of correlation strengths.

\begin{figure}
\centering
\includegraphics[width=\columnwidth]{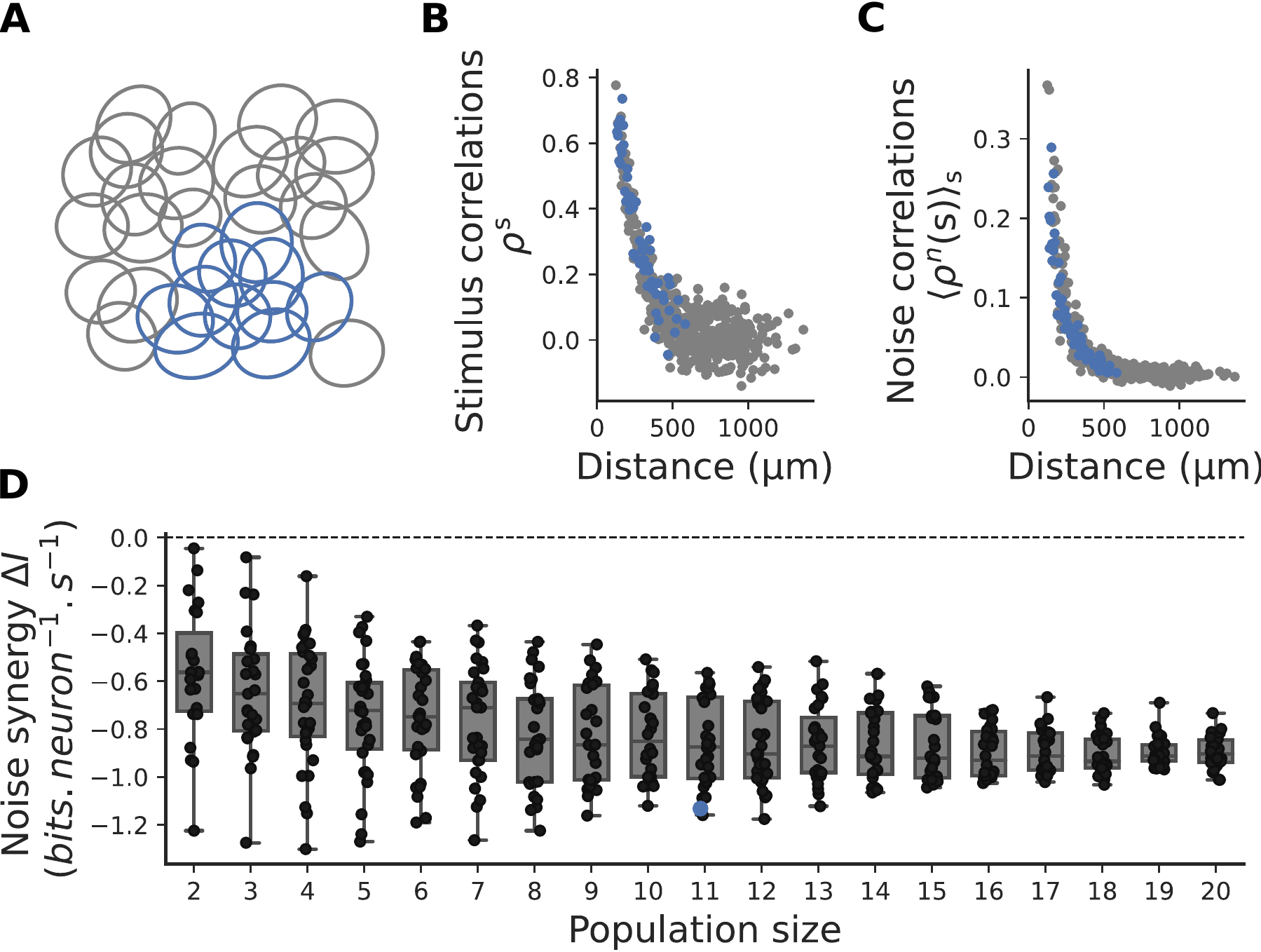}
\caption{\textbf{Application on retinal population response to visual stimulation.}
A) A mosaic of a population of off alpha cells in the rat retina.
B) Stimulus correlation ($\rho^{\rm s}_{ij}$) plotted against the distance between pair of cells stimulated with a white noise movie.
C) Same as (B) but for noise correlations $\<\rho^{\rm n}_{ij}({\bf s})\>_{\bf s}$. 
D) Noise synergy for subset populations of nearby cells. 
Each boxplot corresponds to the noise synergy of many subgroup of ganglion cells. 
Only nearby cells are considered.  
}
\label{f:3A}
\end{figure} 

{\em Application on retinal data.}
We applied our formulas to \textit{ex-vivo} multi-electrode array recordings of rat retinal ganglion cells in response to black and white checkerboard stimulation \cite{Deny17,Sorochynskyi21}.
The receptive fields of the cells have a mosaic structure (Fig.~\ref{f:3A}A), so that neuronal responses show strong stimulus correlations between neighbors, which decay with the distance between the receptive field centers (Fig.~\ref{f:3A}B).
Due to network effects \cite{Brivanlou98},
nearby cells also show strong noise correlations that decay with distance on a similar length scale (Fig.~\ref{f:3A}C).

We computed the noise synergy using our re-summed approximation
\eqref{eq:MI_dep_resumation} for many subgroups of nearby cells of
different sizes (Fig.~\ref{f:3A}D). In this case it is not possible to
estimate mutual informations exactly because of limited data, making
it a good test case for the usefulness of our analytical formulas. To
correct for the bias stemming from noise in estimating
correlations, we subtracted the value obtained after shuffling
individual cell activities across repetitions.
We observe that noise correlations impede information
transmission, by the order of 1 bit per neuron per second, for a total
information of around 10 bits per neuron per second. It should
be stressed however that this result is specific to the white-noise stimulus
statistics considered here, and may not be a general feature of retinal processing, as
other stimulus statistics would change both the nature of stimulus
correlations and the input-output relationship as the network adapts.

\begin{figure}
\centering
\includegraphics[width=\columnwidth]{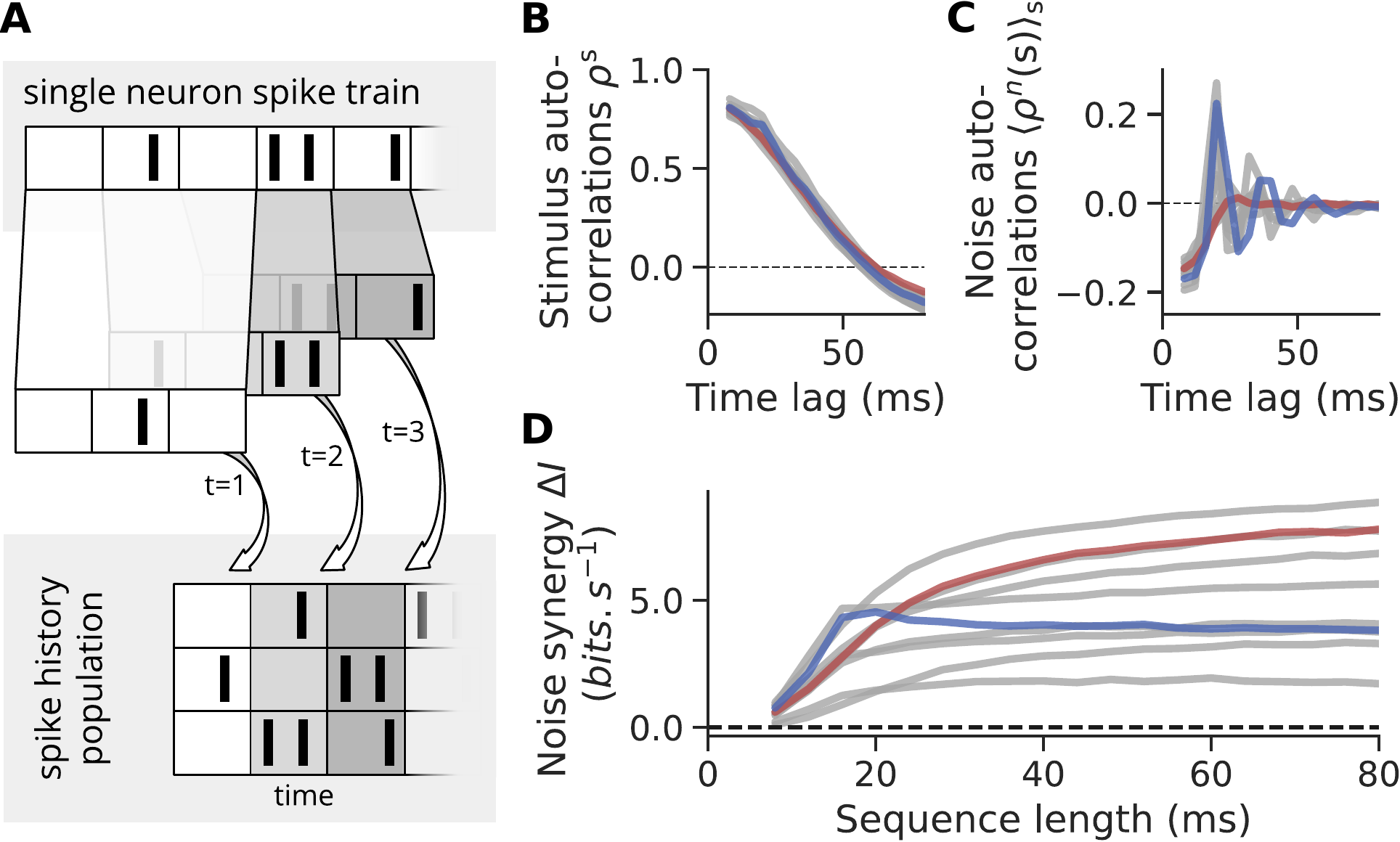}
\caption{\textbf{Application on retinal temporal response to visual stimulation.}
A) We build a pseudo population of neurons to describe the spiking history of single neurons.
B) Stimulus autocorrelations for different cells responding to a white noise stimulation.
Red and blue lines correspond to two example cells.
C) Same as (B) but for noise autocorrelations.
D) Noise synergy for different cells plotted against increasing temporal integration length. 
}
\label{f:3B}
\end{figure}

We also used our method to study the effect of spiking memory in
single neurons, by treating the spike activity of the same neuron in $N$ consecutive time
bins as our activity vector $(n_1,\ldots,n_N)$ (treating time bins as
we treated individual neurons previously, see
Fig.~\ref{f:3B}A). 
Stimulus temporal auto-correlations are positive for about 50ms
(Fig.~\ref{f:3B}B), then become negative and go to zero for longer times (not shown).
Noise temporal correlations are driven by refractoriness, which suppresses activity immediately following a spike, and by burstiness, which induces rippling effects up to 50 ms (Fig.~\ref{f:3B}C).
We find that these
correlations improve information transmission by up to 8 bits per
second (Fig.~\ref{f:3B}D), almost doubling it for some cells. This suggests that information is encoded
not just in the average spike rate, but also through the control of
inter-spike timing, consistent with previous findings \cite{Nemenman08,Ferrari18a,Botella18}.

{\em Discussion.}
Despite being based on a small-correlation expansion, our analytical predictions, especially \eqref{eq:MI_dep_resumation}, work well even in the presence of strong correlations, which are ubiquitous in neuroscience \cite{Brivanlou98, Schneidman06,Moreno-bote14}. We showed how our results can be applied to same-time correlations between neurons, or to neuron autocorrelations, and they can readily be used on general spatial-temporal correlations.

Our work shares some connections with previous efforts to estimate or interpret information in population codes \cite{Brunel98,Panzeri99,Sompolinsky01,Pola03,Hu14,Wei16,Azeredo21}.
Ref. \cite{Pola03} proposes decompositions of the mutual information with different interpretations, but does not provide ways to estimate it.
Refs. \cite{Brunel98} and \cite{Hu14,Wei16} are mostly based on the Fisher information, which in some limit can be related to the mutual information.
While the first term of our simpler expression \eqref{eq:noiseSyn} recovers one of their main results---the so-called sign-rule---second and higher order terms in the noise correlation parameter provide important corrections when correlations are high, as can be seen from deviations from the initial slope in Fig.~\ref{f:2}B.
In \cite{Panzeri99} the authors developed a small time-bin expansion of the mutual information.
Expanding their results for small correlations (and further assuming Poisson distributed spike counts, see App.~D\SI) gives back our second-order expression \eqref{eq:MI_dep_2cd}. Our method however does not need to assume small time bins, and still works well for large correlations.
Ref. \cite{Sompolinsky01} provides estimate of the mutual information when the neuronal responses are correlated but have only small fluctuations around a large mean activity, which is not appropriate for small time bins or for low spike rates as in the retina.

Our results are based on the small correlation expansion developed in \cite{Sessak09}.
In order to apply this theoretical tool, we assumed that both the stimulus-conditioned and the marginal responses follow a pairwise maximum entropy distribution.
These models are characterized by many unknown parameters that in principle need to be inferred from data.
However the final expressions for the mutual information contain only quantities that can be directly estimated from data, without needing any inference.
This makes our approximations ready and easy to use, without requiring much computational efforts. We showed that it works well even when the data was generated with a very different model.
Maximum entropy distributions are actually a series of approximations which, just like Taylor expansions, can be refined by adding higher-order correlations. A future direction could be to compute corrective terms to the mutual information corresponding to third- and higher-order correlation functions, rather than just pairwise correlations as we did in this work. At the same time, the pairwise approximation has proven very accurate for both marginal \cite{Schneidman06,Ganmor11a,Tkacik14,Ganmor15,Ferrari17a} and conditional \cite{Shimazaki12,Granot-Atedgi13,Ferrari18b,Delamare22} responses of populations of neurons, and is only expected to break down for very large densely correlated populations \cite{Roudi09}. We thus expect our results to be applicable to a wide array of neuronal contexts.

\section*{Acknowledgments}
We thank Tobias K\"{u}hn for useful discussions. 
This work was supported by the Agence Nationale de la Recherche (ANR-21-CE37-0024 NatNetNoise), by LabEx LIFESENSES (ANR-10-LABX-65), by IHU FOReSIGHT (ANR-18-IAHU-01), by Sorbonne Université with the Emergence program (CrInforNet), and by Sorbonne Center for Artificial Intelligence - Sorbonne University - IDEX SUPER 11-IDEX-0004.

\onecolumngrid

\appendix


\renewcommand{\thefigure}{S\arabic{figure}}
\setcounter{figure}{0}

\section{Second order approximation}

\subsection{Entropy}
Following \cite{Sessak09} we can expand at small correlations the
entropy of a population of $N$ binary neurons $\mathbf{n} = \left(
  n_1...,n_N \right)$ (taking values $0$ or $1$). Let $\mu_i$ be the
mean of neuron $i$ and $\rho_{i j}$ the Pearson correlation of neurons
$i$ and $j$. The entropy can be expanded as a function of the small
correlation parameter as:
\begin{equation} \label{calc_H}
\begin{split}
    H\left[ \bf n \right] = & H_{\rm 0th}\left[ \bf n \right] + H_{\rm  1st}\left[ \bf n \right] + H_{\rm  2nd}\left[ \bf n \right]+\ldots
\end{split}
\end{equation}
The first term of this expansion $H_{\rm 0th}$ is the entropy of the
neurons in the total absence of correlations and is thus the sum of
the entropies of the single neurons. Adapting the calculation in \cite{Sessak09} to binary neurons taking values 0 or 1 yields:
\begin{equation} \label{calc_H0}
\begin{split}
    H_{\rm 0th}\left[ \bf n \right] = &-\sum_i (1-\mu_i)\log{(1-\mu_i)} + \mu_i\log{(\mu_i)}.
\end{split}
\end{equation}
The first order contribution in the correlation $H_{\rm 1st}$ vanishes and the second order contribution of the correlations to the entropy is given by \cite{Sessak09}:
\begin{equation} \label{calc_H2}
\begin{split}
    H_{\rm 2nd}\left[ \bf n \right] = &-\frac{1}{2}\sum_{i<j}\rho_{i j}^2.
\end{split}
\end{equation}

\subsection{Mutual Information}
We can perform this expansion for the entropy of the marginal distribution $H\left[ \bf n \right]$ as well as for the entropies of the conditional distributions $H\left[ \bf n | \mathbf{s} \right]$, and hence calculate the mutual information to the second order in the correlations:
\begin{equation} \label{calc_I2cdorder}
\begin{split}
    I = I_{\rm 0th}+I_{\rm 2nd},
\end{split}
\end{equation}
where $I_{\rm 0th}$ is the sum of information carried by the neurons individually $I_{\rm 0th}= \sum_i\left[ H[{n_i}] -\<H[n_i|{\bf s}]\>_{\bf s}\right]$:
\begin{equation} \label{calc_I0th}
\begin{split}
    I_{\rm 0th} = &-\sum_i (1-\mu_i)\log{(1-\mu_i)} + \mu_i\log{(\mu_i)} \\
    &+ \left<\sum_i (1-\mu_i(\mathbf{s}))\log{(1-\mu_i(\mathbf{s}))} + \mu_i(\mathbf{s})\log{(\mu_i(\mathbf{s}))}\right>_{\mathbf{s}},
\end{split}
\end{equation}
and where the first non-zero contribution from the pairwise correlations in the response is $I_{\rm 2nd}$:
\begin{equation} \label{calc_I2cd}
\begin{split}
    I_{\rm 2nd} = & -\frac{1}{2} \sum_{i<j} \left( {\rho^{\mathrm{tot}}_{i j}}^2-\left<{\rho_{i j}^{\mathrm{n}}(\mathbf{s})}^2\right>_{\mathbf{s}} \right).
\end{split}
\end{equation}

\subsection{Noise synergy}
Correlations in the marginal response $\rho_{i j}^\text{tot} = r_{i j}^\text{s} + r_{i j}^\text{n}$ boil down to their stimulus contribution in the conditionally independent case $\rho_{i j}^\text{tot, CI} = r_{i j}^\text{s}$. Thus the noise synergy $\Delta I \equiv I - I_\text{CI}$ is given at second oder by:

\begin{align} \label{calc_deltaI2cd}
\Delta I \approx \sum_{i<j} \left[ -r_{i j}^{\mathrm{n}}r_{i j}^{\mathrm{s}} + \frac{1}{2}\left(\left<{\rho_{i j}^{\mathrm{n}}(\mathbf{s})}^2\right>_{\mathbf{s}} - {r_{i j}^{\mathrm{n}}}^2\right) \right].
\end{align}

In general the second term under the sum in the noise synergy, $\Delta I_{i j}^{\rm quad} = \frac{1}{2}\left(\langle{\rho_{i j}^{\mathrm{n}}({\bf s})}^2\rangle_{\bf s} - {r_{i j}^{\mathrm{n}}}^2\right)$ can be positive or negative depending on the level of correlation between ${\rho_{i j}^{\mathrm{n}}({\bf s})}$ and $\sqrt{C_{i i}({\bf s})C_{j j}({\bf s})}$, as $r_{i j}^{\mathrm{n}} = \langle \rho_{i j}^{\mathrm{n}}({\bf s}) \sqrt{C_{i i}({\bf s})C_{j j}({\bf s})} \rangle_{\bf s}/\sqrt{C_{i i} C_{j j}}$. However, if we assume noise correlations are independent from the stimulus $\rho_{i j}^{\mathrm{n}}(\mathbf{s}) = \rho_{i j}^{\mathrm{n}}$, we can show that $\Delta I_{i j}^{\rm quad} = \frac{1}{2}\left({\rho_{i j}^{\mathrm{n}}}^2 - {r_{i j}^{\mathrm{n}}}^2\right)$ is non-negative. First, the formulation of ${r_{i j}^{\mathrm{n}}}$ in terms of ${\rho_{i j}^{\mathrm{n}}}$ becomes:
\begin{equation}
{r_{i j}^{\mathrm{n}}} = {\rho_{i j}^{\mathrm{n}}}\frac{\left<\sqrt{C_{ii}^{\rm n}(\mathbf{s})C_{jj}^{\rm n}(\mathbf{s})}\right>_{\rm s}}{\sqrt{C_{ii}C_{jj}}},
\end{equation}
which then gives:
\begin{equation}
\Delta I_{i j}^{\rm quad} = \frac{1}{2} {\rho_{i j}^{\mathrm{n}}}^2 \left( 1 - \frac{\left<\sqrt{C_{ii}^{\rm n}(\mathbf{s})C_{jj}^{\rm n}(\mathbf{s})}\right>_{\rm s}^2}{C_{ii}C_{jj}} \right).
\end{equation}
According to the Cauchy-Schwartz inequality we have that $\left<\sqrt{C_{ii}^{\rm n}(\mathbf{s})C_{jj}^{\rm n}(\mathbf{s})}\right>_{\rm s}^2 \leq \left< C_{ii}^{\rm n}(\mathbf{s})\right>_{\rm s} \left< C_{jj}^{\rm n}(\mathbf{s})\right>_{\rm s}$. Besides, from the law of total variance we have that $C_{ii} = C_{ii}^{\rm s} + \left< C_{ii}^{\rm n}(\mathbf{s})\right>_{\rm s}$, thus $\left< C_{ii}^{\rm n}(\mathbf{s})\right>_{\rm s} \leq C_{ii} $ and finally ${\left<\sqrt{C_{ii}^{\rm n}(\mathbf{s})C_{jj}^{\rm n}(\mathbf{s})}\right>_{\rm s}^2} \leq {C_{ii}C_{jj}}$. Altogether this gives that $\Delta I_{i j}^{\rm quad} \geq 0$.

\section{Resummed expansion}

\subsection{Entropy}
In \cite{Sessak09} it is shown that some of the terms in the small-correlation expansion of the couplings can be resummed to yield a better approximation. We can proceed in the exact same way for the entropy and resum some of the diagrams in the small-correlation expansion. All the terms of the second order approximation above are contained in the resummed expansion we detail here. Note that this resummed expansion is equivalent to a cluster expansion truncated to second order, with mean-field reference entropy \cite{Cocco12}. It reads:
\begin{align}\label{eq:calc_H_resumation}
H\left[ \bf n \right] &\approx H_{\rm 0th}\left[ \bf n \right] + H_{\rm  pairs}\left[ \bf n \right]  + H_{\mathcal{G}}\left[ \bf n \right] - H_{\rm dbl}\left[ \bf n \right].
\end{align}
The single site contribution $H_{\rm 0th}\left[ \bf n \right]$ is the same as above, and $H_{\rm  pairs}\left[ \bf n \right]$ corresponds to the entropy gain of all pairs in the population taken independently compared to the single site contribution. Interestingly, taking the sum of these two first contributions amounts to making an independent pair approximation, which would be exact in the case of a tree-like network topology. $H_{\rm  pairs}\left[ \bf n \right]$ is a sum over all pairs of neurons in the population:
\begin{align}\label{eq:calc_H_pairs}
H_{\rm  pairs}\left[ \bf n \right] = \sum_{i<j}H[n_i, n_j],
\end{align}
with $H[n_i, n_j]$ the entropy gain of pair $(i j)$ compared to the single neurons case:
\begin{equation}\label{eq:cal_H_i j}
\begin{split} 
    H[n_i, n_j] = & -(C_{i j}+\mu_i \mu_j)\log{(1+\frac{C_{i j}}{\mu_i \mu_j})} \\
    & + (C_{i j} + \mu_i (\mu_j - 1))\log{(1+\frac{C_{i j}}{\mu_i(\mu_j-1)})} \\
    & + (C_{i j} + \mu_j(\mu_i - 1))\log{(1+\frac{C_{i j}}{\mu_j(\mu_i-1)})} \\
    & - (C_{i j} + (1-\mu_i)(1-\mu_j))\log{(1+\frac{C_{i j}}{(1-\mu_i)(1-\mu_j)})}.
\end{split}
\end{equation}
The second resummed term of this expansion $H_{\mathcal{G}}\left[ \bf n \right]$ corresponds to the contribution of interactions to the entropy in the mean-field approximation. It contains the resummation of all loop diagrams in the expansion and amounts to assuming the entropic contribution of pairwise correlations is Gaussian. Noting $\rho$ the correlation matrix of $\bf n$ we get:

\begin{align}\label{eq:calc_H_gauss}
H_{\mathcal{G}}\left[ \bf n \right] = \frac{1}{2}\log\left(\left| \bf \rho \right|\right).
\end{align}

Finally there are some terms in the expansion that are resummed both in $H_{\rm  pairs}\left[ \bf n \right]$ and $H_{\mathcal{G}}\left[ \bf n \right]$, therefore we need to substract them once from the expansion through $H_{\rm  dbl}\left[ \bf n \right]$. The double counted terms simply correspond to the Gaussian approximation applied to each pair of neurons:
\begin{align}\label{eq:calc_H_dbl}
H_{\rm dbl}\left[ \bf n \right] = \frac{1}{2}\sum_{i<j}\log\left(1-\rho_{i j}^2\right).
\end{align}

\subsection{Mutual Information}
Applying the resummed approximation to the marginal and conditional responses results in a resummed approximation for the mutual information:

\begin{align}\label{eq:cal_I_resumation}
I &\approx I_{\rm 0th} + I_{\rm  pairs}  + I_{\mathcal{G}} - I_{\rm dbl}.
\end{align}

With $I_{\rm  pairs}$ the contribution of pairwise correlations to the mutual information in the independent pairs approximation. If we write $C_{i j}\equiv\mathrm{Cov}(n_i,n_j)$ the total covariance across stimuli and $C_{i j}^{\rm n}({\bf s})\equiv\mathrm{Cov}(n_i,n_j|{\bf s})$ the covariance at a given stimulus $\bf s$:
\begin{equation} \label{calc_Ipairs}
\begin{split}
I_{\rm pairs} = & \sum_{i< j} \biggl[-(C_{i j}+\mu_i \mu_j)\log{(1+\frac{C_{i j}}{\mu_i \mu_j})} \\
    & + (C_{i j} + \mu_i (\mu_j - 1))\log{(1+\frac{C_{i j}}{\mu_i(\mu_j-1)})} \\
    & + (C_{i j} + \mu_j(\mu_i - 1))\log{(1+\frac{C_{i j}}{\mu_j(\mu_i-1)})} \\
    & - (C_{i j} + (1-\mu_i)(1-\mu_j))\log{(1+\frac{C_{i j}}{(1-\mu_i)(1-\mu_j)})} \\
    & -\biggl<-(C_{i j}^{\mathrm{n}}(\mathbf{s})+\mu_i(\mathbf{s}) \mu_j(\mathbf{s}))\log{(1+\frac{C_{i j}^{\mathrm{n}}(\mathbf{s})}{\mu_i(\mathbf{s}) \mu_j(\mathbf{s})})} \\
    & + (C_{i j}^{\mathrm{n}}(\mathbf{s}) + \mu_i(\mathbf{s}) (\mu_j(\mathbf{s}) - 1))\log{(1+\frac{C_{i j}^{\mathrm{n}}(\mathbf{s})}{\mu_i(\mathbf{s})(\mu_j(\mathbf{s})-1)})} \\
    & + (C_{i j}^{\mathrm{n}}(\mathbf{s}) + \mu_j(\mathbf{s})(\mu_i(\mathbf{s}) - 1))\log{(1+\frac{C_{i j}^{\mathrm{n}}(\mathbf{s})}{\mu_j(\mathbf{s})(\mu_i(\mathbf{s})-1)})} \\
    & -(C_{i j}^{\mathrm{n}}(\mathbf{s}) + (1-\mu_i(\mathbf{s}))(1-\mu_j(\mathbf{s})))\log{(1+\frac{C_{i j}^{\mathrm{n}}(\mathbf{s})}{(1-\mu_i(\mathbf{s}))(1-\mu_j(\mathbf{s}))})}\biggr> _{\mathbf{s}}\biggr].
\end{split}
\end{equation}
If we denote by $\rho^{\rm tot}$ the total correlation matrix across stimuli and $\rho^{\rm n}\left( \mathbf{s}\right)$ the correlation matrix at given stimulus $\bf s$, the Gaussian (i.e. mean-field) contribution of correlations to the mutual information takes the simple form:
\begin{align} \label{eq:cal_Igauss}
I_{\mathcal{G}} = \frac{1}{2}\log{\left( |\rho^{\mathrm{tot}}| \right)} - \frac{1}{2}\left<\log{\left( |\rho^{\mathrm{n}}(\mathbf{s})|\right)}\right>_{\mathbf{s}},
\end{align}
while the double counting correction becomes:

\begin{align} \label{eq:cal_Idbl}
I_{\rm dbl} = \frac{1}{2}\sum_{i<j}\biggl[\log\left(1-{\rho_{i j}^{\rm tot}}^2\right) -  \left<\log\left(1-{\rho_{i j}^{\rm n}}\left( \bf s\right)^2\right)\right>_{\mathbf{s}}\biggr].
\end{align}

\subsection{Noise synergy}
Likewise we can write a resummed approximation of the noise synergy:

\begin{align}\label{eq:cal_deltaI_resumation}
\Delta I &\approx \Delta I_{\rm  pairs}  + \Delta I_{\mathcal{G}} - \Delta I_{\rm dbl}.
\end{align}

Noting that the total covariance across stimuli can be decomposed in terms of noise covariance and stimulus covariance $C_{i j} = C_{i j}^{\rm s} + C_{i j}^{\rm n}$, with $C_{i j}^{\rm n} = \langle C_{i j}^{\rm n}(\mathbf{s}) \rangle_{\rm s}$, the independent pairs approximation of the noise synergy $\Delta I_{\rm  pairs}$ would be given by:

\begin{equation} \label{cal_deltaI_pairs}
\begin{split}
\Delta I_{\rm pairs} = & \sum_{i<j} \biggl[- C_{i j}^{\rm n} \log(1 + \frac{C_{i j}}{(\mu_i(\mu_j-1)+C_{i j})(\mu_j(\mu_i-1)+C_{i j})}) \\
    & - (C_{i j}^{\mathrm{s}} + \mu_i \mu_j) \log(1+ \frac{C_{i j}^{\rm n}}{\mu_i \mu_j + C_{i j}^{\mathrm{s}}}) \\
    & + (C_{i j}^{\mathrm{s}} + \mu_i(\mu_j-1)) \log(1+ \frac{C_{i j}^{\rm n}}{\mu_i(\mu_j-1) + C_{i j}^{\mathrm{s}}}) \\
    & + (C_{i j}^{\mathrm{s}} + \mu_j(\mu_i-1)) \log(1+ \frac{C_{i j}^{\rm n}}{\mu_j(\mu_i-1) + C_{i j}^{\mathrm{s}}}) \\
    & - (C_{i j}^{\mathrm{s}} + (1-\mu_i)(1-\mu_j)) \log(1+ \frac{C_{i j}^{\rm n}}{(1-\mu_i)(1-\mu_j) + C_{i j}^{\mathrm{s}}}) \\
    & -\biggl<-(C_{i j}^{\mathrm{n}}(\mathbf{s})+\mu_i(\mathbf{s}) \mu_j(\mathbf{s}))\log{(1+\frac{C_{i j}^{\mathrm{n}}(\mathbf{s})}{\mu_i(\mathbf{s}) \mu_j(\mathbf{s})})} \\
    & + (C_{i j}^{\mathrm{n}}(\mathbf{s}) + \mu_i(\mathbf{s}) (\mu_j(\mathbf{s}) - 1))\log{(1+\frac{C_{i j}^{\mathrm{n}}(\mathbf{s})}{\mu_i(\mathbf{s})(\mu_j(\mathbf{s})-1)})} \\
    & + (C_{i j}^{\mathrm{n}}(\mathbf{s}) + \mu_j(\mathbf{s})(\mu_i(\mathbf{s}) - 1))\log{(1+\frac{C_{i j}^{\mathrm{n}}(\mathbf{s})}{\mu_j(\mathbf{s})(\mu_i(\mathbf{s})-1)})} \\
    & -(C_{i j}^{\mathrm{n}}(\mathbf{s}) + (1-\mu_i(\mathbf{s}))(1-\mu_j(\mathbf{s})))\log{(1+\frac{C_{i j}^{\mathrm{n}}(\mathbf{s})}{(1-\mu_i(\mathbf{s}))(1-\mu_j(\mathbf{s}))})}\biggr> _{\mathrm{s}}\biggr].
\end{split}
\end{equation}

The Gaussian contribution to the noise synergy takes again a simple form:
\begin{equation} \label{cal_deltaI_gauss}
\begin{split}
\Delta I_{\mathcal{G}} &= \frac{1}{2}\log \left( \frac{|\mathbf{\rho}^{\mathrm{tot}}|}{|\mathbf{\rho}^{\mathrm{tot, CI}}|} \right) - \frac{1}{2}\left <\log \left( |\mathbf{\rho}^{\mathrm{n}}(\mathbf{s})| \right)\right >_\mathbf{s},
\end{split}
\end{equation}
where we recall that correlations in the marginal response can be expressed as the sum of stimulus and noise contributions $\mathbf{\rho}^{\mathrm{tot}} = r^\text{s} + r^\text{n}$. In the conditionally independent case we have $\mathbf{\rho}^{\mathrm{tot, CI}} = r^\text{s} + \nu^\text{n}$ with $\nu^\text{n}$ the diagonal matrix containing the diagonal elements of $r^\text{n}$.

Finally, the double counting correction to the noise synergy is given by:
\begin{equation} \label{cal_deltaI_double}
\begin{split}
\Delta I^{dbl} = &\frac{1}{2}\sum_{i<j}\biggl[\log{\left( \frac{1-{\rho^{\rm tot}_{i j}}^2}{1-{r^{\mathrm{s}}_{i j}}^2} \right)} - \left< \log{\left( 1-{\rho_{i j}^{\mathrm{n}}(\mathbf{s})}^2 \right)} \right>_{\mathrm{s}}\biggr].
\end{split}
\end{equation}

\section{Generalized linear model simulations}
The model used to generate the synthetic data is a Generalized Linear Model \cite{Pillow08} with sigmoidal nonlinearity. The number of spikes emitted by cell $i$ in time bin $t$ of size $dt=1$\,ms follows a Bernoulli distribution with mean $\lambda_i\left(t\right)$, given by:
\begin{align} \label{eq:GLM_def}
&\lambda_i\left(t\right) = \left(1+\exp{\left(- h_i\left(t\right) \right)}\right)^{-1},
&h_i\left(t\right) = h_i^{\rm bias} + h_i^{\rm stim}\left(t\right) + h_i^{\rm int}\left(t\right) + h_i^{\rm corr}\left(t\right),
\end{align}
where $h_i^{\rm bias}$ sets the baseline firing rate of the cell, $h_i^{\rm stim}\left(t\right)$ accounts for how the stimulus drives the cell's activity, $h_i^{\rm int}\left(t\right)$ accounts for the effect of couplings and self-coupling, while $h_i^{\rm corr}\left(t\right)$ is here to correct for the contribution of the neuron-neuron couplings to the firing rate.

\begin{figure}[t!]
\centering
\includegraphics[width=\columnwidth]{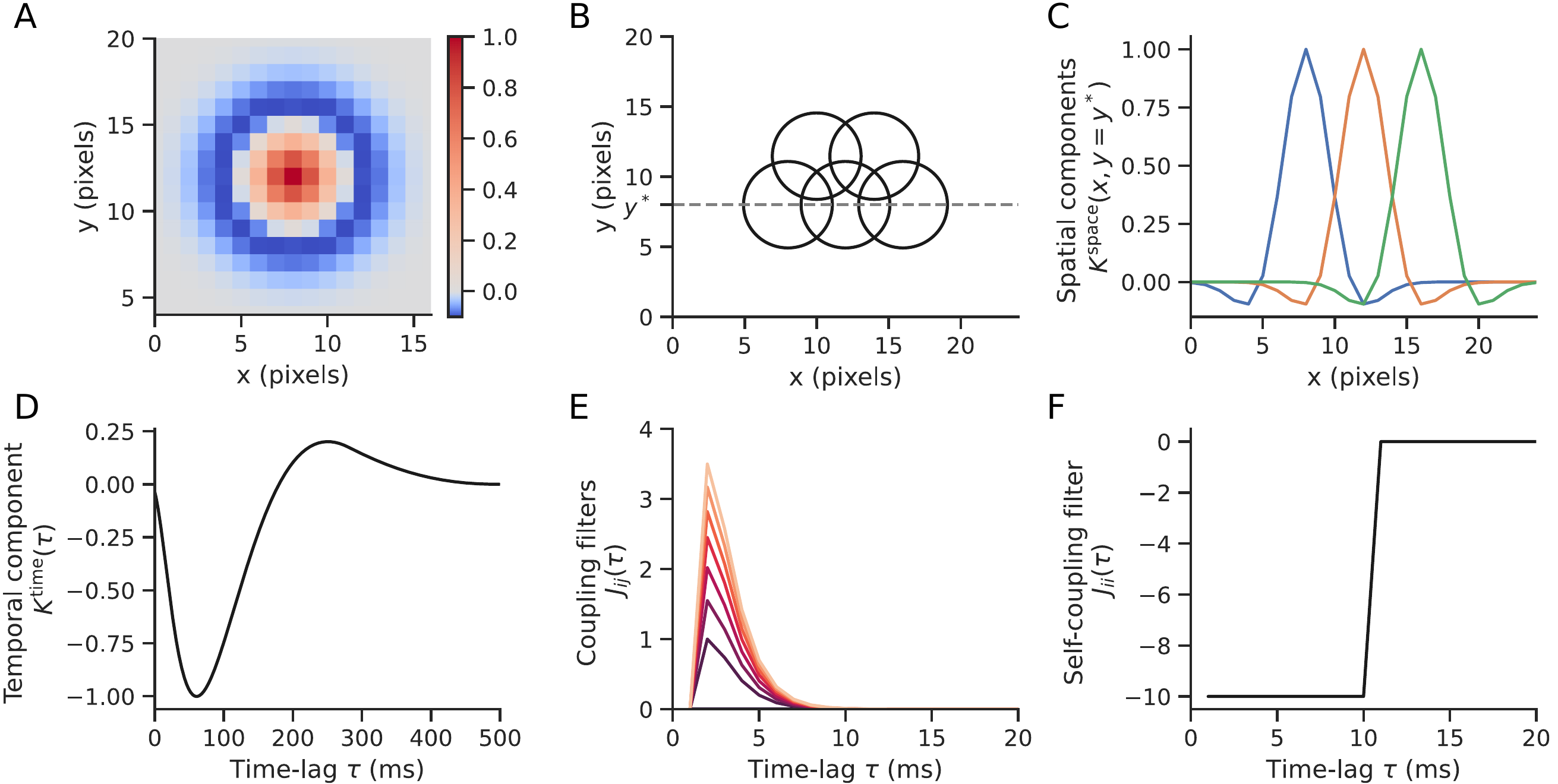}
\caption{\textbf{Parameters of the GLM model.}
  The different parameters of the model were chosen to be biologically plausible and mimick those of retinal ganglion cells.
  A) The spatial component of the spatio-temporal stimulus filter $K^{\rm space}$ is made of a difference of Gaussians Eq. \ref{eq:Kspace}.
  B) The 5 cells are arranged according to a triangular lattice. We represented here the receptive fields of the cells by circles that correspond to the contour where the center and the surround of the spatial components compensate exactly. Only nearest neighbours in this population are coupled by non-zero $J_{ij}$ coupling filters.
  C) Profile view of the spatial components cut in $y=y^*$ on panel B.
  D) The temporal component of the stimulus filter has a biphasic profile and consists of a difference of two raised cosine functions Eq. \ref{eq:Ktime}.
  E) Coupling filters are non zero for nearest neighbours and are defined by Eq. \ref{eq:Jtau}. The increasing coupling amplitude shown here induces increasing noise correlations in the response of the cells.
  F) The self coupling filter, accounting for the effect of the neuron's own spiking history, induced refractory effects over $10$\,ms in the past.}
\label{f:sup1}
\end{figure}

The stimulus $S$ is a movie of dimensions $(N_x, N_y, N_t)$ where $N_x$ and $N_y$ correspond to the two spatial dimensions in pixels, and $N_t$ is the temporal length of the stimulus in number of time bins.
We simulated $5$ cells organized on a triangular lattice as represented in Fig.~\ref{f:sup1}B, spaced by $\xi = 4$ pixels. Here the stimulus consisted of $N_t$ zero-mean, 2D Gaussian frames of size $(N_x, N_y)$ with covariance function $C({\bf u},{\bf v}) = \delta_{{\bf u},{\bf v}} + (1-\delta_{{\bf u},{\bf v}}) c_0 \exp{\left( -\|{\bf u}-{\bf v}\|/\lambda \right)}$, where $\lambda = 2\xi$ and $c_0$ is varied to change the level of stimulus correlation in the response. There was no correlations between frames and they were refreshed at 100Hz.

The cells' stimulus filters, of size $(N_x, N_y, N_t^{\rm stim})$ were built from the product of a temporal and spatial component which were chosen to mimick the properties of natural retinal ganglion cells.
The temporal component $ K^{\rm time}(\tau) $ is illustrated Fig.~\ref{f:sup1}D and consists of a difference of two raised cosine functions:
\begin{equation}
{\rm rc}(\tau, s, c) = 
\begin{cases}
\cos((\log(\tau+s)-c)\pi)/2+1/2 & \text{when} -1\leq\log(\tau+s)-c\leq1 \\ \\
0 & \text{otherwise},
\end{cases}
\end{equation}
such that:
\begin{equation} \label{eq:Ktime}
K^{\rm time}(\tau) = a_1 {\rm rc}(\tau, s, c_1) - a_2 {\rm rc}(\tau, s, c_2),
\end{equation}
with $s=50$, $a_1 = 0.35$, $c_1 = 5.3$, $a_2 = 1.15$ and $c_2 = 4.8$.
The spatial component is made of a difference of Gaussian functions and exhibits a positive center and a negative surround, as shown on Fig.~\ref{f:sup1}A and Fig.~\ref{f:sup1}C. Noting ${\bf u^{\rm c}_i}=(x^{\rm c}_i, y^{\rm c}_i)$ the spatial position of the receptive field center of cell $i$:
\begin{equation} \label{eq:Kspace}
K^{\rm space}_i({\bf u}) = \frac{1}{r-1}\left[r \exp\left(-\frac{\| {\bf u}-{\bf u^{\rm c}_i} \|^2}{2\sigma_{\rm center}}\right) - \exp\left(-\frac{\| {\bf u}-{\bf u^{\rm c}_i} \|^2}{2\sigma_{\rm surround}}\right)\right]
\end{equation}
where $\sigma_{\rm center}=2$ pixel, $\sigma_{\rm surround}=2.1$ pixel, and $r=1.12$. The firing rate variance is fixed through a parameter $\alpha_{\rm s}$ (set to $0.5$ in our synthetic experiments) such that $h_i^{\rm stim}\left(t\right) = \alpha_{\rm s} \cdot z\text{-score}\left(\tilde{h}_i^{\rm stim}\left(t\right)\right)$, where $\tilde{h}_i^{\rm stim}\left(t\right)$ is given by the temporal convolution of stimulus $S$ by the spatio-temporal stimulus filter:
\begin{equation} \label{eq:GLM_hstim}
h_i^{\rm stim}\left(t\right) = \sum_{\tau>0}\sum_{\bf u}K_i^{\rm space}({\bf u})K^{\rm time}(\tau)S({\bf u},t-\tau).
\end{equation}

Likewise, the spiking history contribution of cell $i$ itself as well as that of the other cells in the network are accounted for by linear convolutions of the spiking histories by a set of temporal coupling filters:
\begin{align} \label{eq:GLM_hint}
h_i^{\rm int}\left(t\right) = \sum_{j}\sum_{\tau>0}J_{i j}\left(\tau\right) n_{j}(t-\tau),
\end{align}
where the self-coupling filters $J_{ii}(\tau)$, shown Fig.~\ref{f:sup1}F, are given by:
\begin{equation}
J_{ii}(\tau) = 
\begin{cases}
J_{\rm self}^0 & \text{if} ~ \tau \leq \tau_{\rm refr,} \\ \\
0 & \text{otherwise,}
\end{cases}
\end{equation}
with $J_{\rm self}^0=-10$ and $\tau_{\rm refr}=10\text{ms}$. The neuron-neuron (i.e. $i \neq j$) couplings $J_{ij}$ follow:
\begin{equation} \label{eq:Jtau}
J_{ij}(\tau) =
\begin{cases}
J_{\rm coupl}^0 ~ \tau \exp\left(-\tau\right) & \text{if i and j are nearest neighbours,} \\ \\
0 & \text{otherwise,}
\end{cases}
\end{equation}
where the coupling strength $J_{\rm coupl}^0$ can be varied (as illustrated on Fig.~\ref{f:sup1}E) to change the amount of noise correlations in the response.

\begin{figure}[t!]
\centering
\includegraphics[width=\columnwidth]{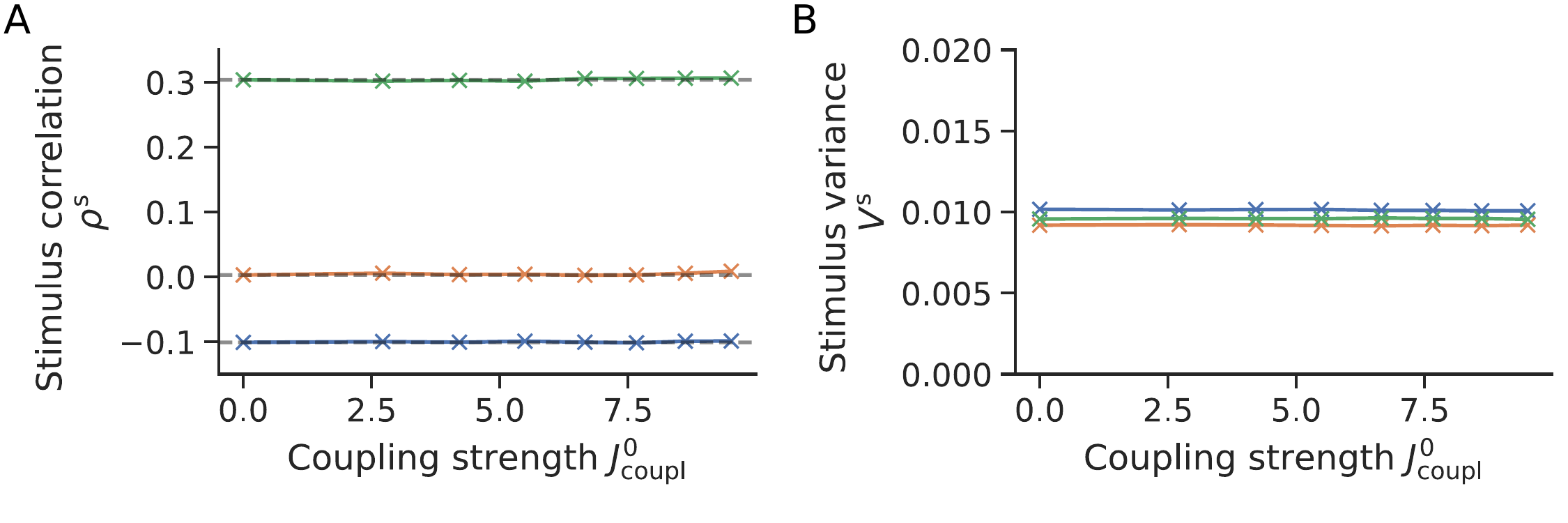}
\caption{\textbf{Results of the iterative inference approach.}
  The different parameters of the model were chosen to be biologically plausible and mimick those of retinal ganglion cells.
  A) For the three test stimuli chosen we see that $\rho^{\rm s}$, the stimulus correlation averaged over all pairs of neighbouring cells, is constant with respect to the coupling strength $J_{\rm coupl}^0.$
  B) Likewise, the stimulus variance averaged across the population $V^{\rm s} = \langle C_{i i}^{\rm s} \rangle_i$ is independent of $J_{\rm coupl}^0$.}
\label{f:sup2}
\end{figure}

In order to vary independently stimulus and noise correlations in the neurons' response we introduced the field $h_i^{\rm corr}$ that corrects for the contribution of neuron-neuron couplings to the firing rates of the cells. This correction is needed because changing the strength of the couplings will not only change the amount of noise correlations in the reponse, but also the firing rates of the cells. This corrective field is computed via an iterative inference approach built upon the 2-step inference method \cite{Mahuas20}. The first step of this iterative procedure is the following: for a given stimulus movie, we simulate $N_{\rm repe}$ times the response of the conditionally-independent neurons (i.e. with $J_{\rm coupl}^0 = 0$). We then fix the strength of neuron-neuron couplings $J_{\rm coupl}^0$ to the desired value so as to induce noise correlations in the response of the cells. From here we approximate the interaction field $\hat{h}_i^{\rm int}$ using the simulated response and the chosen couplings amplitude $J_{\rm coupl}^0$. Finally we infer the corrective field $h_i^{\rm corr}$ on the previously simulated response similarly to the 2-step inference approach by minimzing the following single neuron negative log-likelihood for each neuron in the population:
\begin{equation}
nLL_i = - \sum_{r=1}^{N_{\rm repe}} \sum_{t=1}^{N_{\rm t}} h_{i}^{\rm corr}(t) n_i(t,r) + \log{\left( 1-\hat{\lambda}_i(t,r) \right)},
\end{equation}
where $n_i(t,r)$ denotes the simulated response of neuron $i$ in time bin $t$ and repetition $r$ and where $\hat{\lambda}_i(t,r) = \left(1+\exp{\left(- h_i^{\rm bias} + h_i^{\rm stim}(t) + \hat{h}_i^{\rm int}(t,r) + h_i^{\rm corr}(t) \right)}\right)^{-1}$. The second step of the procedure consists in simulating the response of the cells using the previously chosen coupling amplitude and the infered corrective field $h_i^{\rm corr}\left(t\right)$. We can then re-estimate the interaction field $\hat{h}_i^{\rm int}$ on these simulated data and infer again $h_i^{\rm corr}$ to get a better approximation of the couplings' contribution to the firing rate. This second step is repeated as many times as needed to match the firing rates of the conditionally-dependent model to those of the conditionally-independent model within the desired precision. For the generation of the synthetic data used for testing the approximations we systematically performed 8 successive inference steps. This approach resulted for each given stimuli in constant stimulus correlations Fig.~\ref{f:sup2}A as well as constant stimulus variance Fig.~\ref{f:sup2}B across coupling strengths.

The simulation of the model described above was repeated $N_{\rm repe}=2500$ times for each stimulus movie, then the Mutual Information was computed using the second order and the resummed approximations as well as by the histogram (or ``exact") method. In order to be able to compare the Mutual Information computed via the approximations to that computed via the histogram method, we first need to correct for the effect of sampling bias on these quantities. From \cite{Macke11} the bias in the entropy of a maximum entropy model with $N_{\rm cells}$ and thus $\frac{N_{\rm cells}(N_{\rm cells}+1)}{2}$ constraints, evaluated on $N_{\rm samples}$ can be approximated by $b^{\rm maxent}_{\rm H} = -\frac{N_{\rm cells}(N_{\rm cells}+1)}{4 N_{\rm samples}}$. In the case of the marginal entropy we have $N_{\rm samples} = N_{\rm repe} \times N_{t}$, while in the case of the noise entropy we have $N_{\rm samples} = N_{\rm repe}$. The sampling bias on the Mutual Information of a maximum entropy model is thus given by:
\begin{equation}
b^{\rm maxent}_{\rm I} = \frac{N_{\rm cells}(N_{\rm cells}+1)}{4 N_{\rm repe}}(1-\frac{1}{N_t}),
\end{equation}
which has then to be subtracted from the raw results of the 2nd order and the resummed approximations.
To evaluate the bias on the entropies evaluated via the histogram method we used a shuffling approach similar to \cite{Montemurro2007}: we estimate the bias on the entropy $H[{\bf n}]$ as $b^{\rm exact}_{\rm H} = H^{\rm shuffle}[{\bf n}]-(H^{\rm 0th}[{\bf n}] - b^{0th})$ where $H^{\rm shuffle}[{\bf n}]$ is the entropy computed on the data shuffled so that correlations between cells are destroyed, and where $H^{\rm 0th}[{\bf n}]$ is the single site entropy contribution given by Eq.\ref{calc_H0}. $H^{\rm 0th}[{\bf n}]$ is also biased, so we need to correct it by $b^{\rm 0th}_{H}=-\frac{N_{\rm cells}}{2N_{\rm samples}}$ (as here we have only one constraint per neuron). Applying this to the marginal and conditional entropies gives the following bias for the Mutual Information computed via the exact method:
\begin{equation}
b^{\rm exact}_{\rm I} = \left(H^{\rm shuffle}[{\bf n}] - \langle H^{\rm shuffle}[{\bf n | s}] \rangle_{\rm s} \right) - \left( H^{\rm 0th}[{\bf n}] - \langle H^{\rm 0th}[{\bf n | s}] \rangle_{\rm s} \right) + \frac{N_{\rm cells}}{2 N_{\rm repe}}(1-\frac{1}{N_t}),
\end{equation}
which has to be subtracted from the raw result of the histogram method.

\section{Link to Panzeri {\em et al.} 1999}

\subsection{Small time bin expansion}
Panzeri et al. \cite{Panzeri99} introduced a small time bin $t$ expansion of the mutual information:
\begin{equation} \label{I_Panzeri_2cd_order}
    I = t I^t + \frac{t^2}{2} I^{tt} + ...
\end{equation}
They quantify noise correlations and stimulus correlations in terms of $n_i(\mathbf{s})$ the spike count variable at fixed stimulus $s$, $\mu_i(\mathbf{s}) = \overline{n_i(\mathbf{s})}$ its average over the noise, $\mu_i = \left<\mu_i(\mathbf{s})\right>_{\bf s}$ the average firing rate across stimuli and use the according rates $n_i(\mathbf{s})/t$, $\mu_i(\mathbf{s})/t$ and $\mu_i/t$ in the expression of the mutual information. They introduce the ``noise correlation density":
\begin{equation}
    \gamma_{i j}(\mathbf{s}) =
    \begin{cases}
      \frac{\overline{n_i(\mathbf{s})n_j(\mathbf{s})}}{\mu_i(\mathbf{s})\mu_j(\mathbf{s})}-1 & \text{if}\ i\neq j \\ \\
      \frac{\overline{{n_i(\mathbf{s})}^2}-\mu_i(\mathbf{s})}{{\mu_i(\mathbf{s})}^2}-1 & \text{if}\ i = j,
    \end{cases}
\end{equation}
and the ``stimulus correlation density":
\begin{equation}
    \nu_{i j} = \frac{\left<\mu_i(\mathbf{s})\mu_j(\mathbf{s})\right>_{s}}{\mu_i\mu_j}-1.
\end{equation}
The first order contribution (sum over the single cell contributions) to the mutual information in the small time bin expansion is:
\begin{equation} \label{It}
t I^{t}=\sum_{i}\left\langle\mu_{i}(\mathbf{s}) \ln \frac{\mu_{i}(\mathbf{s})}{\mu_{i}}\right\rangle_{s},
\end{equation}
The second order contribution contains the correlations contributions:

\begin{equation}\label{Itt}
\frac{t^2}{2}I^{t t}= \frac{t^2}{2}\sum_{i j} I^{t t, {\rm (1)}}_{i j} + I^{t t, {\rm (2)}}_{i j} + I^{t t, {\rm (3)}}_{i j},
\end{equation}
with
\begin{align}
I^{t t, {\rm (1)}}_{i j} = & \frac{1}{t^2}\mu_{i}\mu_{j}\left[\nu_{i j}+\left(1+\nu_{i j}\right) \ln \left(\frac{1}{1+\nu_{i j}}\right)\right], \\
I^{t t, {\rm (2)}}_{i j} = & \frac{1}{t^2}\left\langle\mu_{i}(\mathbf{s}) \mu_{j}(\mathbf{s}) \gamma_{i j}(\mathbf{s})\right\rangle_{s} \ln\left(\frac{1}{1+\nu_{i j}}\right), \\
I^{t t, {\rm (3)}}_{i j} = & \frac{1}{t^2}\bigg\langle\mu_{i}(\mathbf{s}) \mu_{j}(\mathbf{s})\left(1+\gamma_{i j}(\mathbf{s})\right) \times \ln\left(\frac{\left(1+\gamma_{i j}(\mathbf{s})\right)\left\langle\mu_{i}\left(\mathbf{s}^{\prime}\right) \mu_{j}\left(\mathbf{s}^{\prime}\right)\right\rangle_{s^{\prime}}}{\left\langle\mu_{i}\left(\mathbf{s}^{\prime}\right) \mu_{j}\left(\mathbf{s}^{\prime}\right)\left(1+\gamma_{i j}\left(\mathbf{s}^{\prime}\right)\right)\right\rangle_{s^{\prime}}}\right)\bigg\rangle_{s}.
\end{align}
The three contributions in $I^{t t}$ render the effects of stimulus and noise correlations as well as interactions thereof \cite{Panzeri99}: $I^{t t, {\rm (1)}}_{i j}$ contains the effect of signal correlations, while $I^{t t, {\rm (2)}}_{i j}$ accounts for how (stimulus independent) noise correlations interact with stimulus correlations and affect information, and $I^{t t, {\rm (3)}}_{i j}$ contains information carried by the stimulus-dependency of noise correlations.

\subsection{Link to the small correlation expansion}
The single site contribution of the small time bin approximation $t I_{\rm t}$ amounts to assuming information is conveyed only by spikes rather than by spikes and silences together. To illustrate this, we can rewrite the expression obtained previously for $I_{\rm 0th}$ by regrouping the noise entropy and marginal entropy contributions separately for spikes on one hand, and silences on the other hand:
\begin{equation} \label{calc_I0th}
\begin{split}
    I_{\rm 0th} = \sum_i \left< \mu_i(\mathbf{s})\log{\left(\frac{\mu_i(\mathbf{s})}{\mu_i}\right)}\right>_{\mathbf{s}} + \sum_i \left< (1-\mu_i(\mathbf{s}))\log{\left(\frac{1-\mu_i(\mathbf{s})}{1-\mu_i}\right)}\right>_{\mathbf{s}}.
\end{split}
\end{equation}
We see the first term in this rewriting of $I_{\rm 0th}$ corresponds to $t I^{t}$. Further corrections to the single site information $I_{\rm t}$ are found in the diagonal terms under the sum in $I^{t t}$. In the small rates (i.e. Poisson) limit however, this correction is simplified as $\gamma_{ii}(\mathbf{s})$ vanishes for $C_{i i}({\bf s}) = \mu_i({\bf s})$.
The single site correction coming from $I^{t t}$ is therefore given by:
\begin{equation} \label{calc_diag_Itt}
\frac{t^2}{2} \sum_{i} I^{t t, {\rm (1)}}_{ii} = \frac{1}{2} \sum_{i}{\mu_{i}}^2\left[\nu_{i i}+\left(1+\nu_{i i}\right) \ln \left(\frac{1}{1+\nu_{i i}}\right)\right].
\end{equation}
The out of diagonal terms ($i \neq j $) in $I^{t t}$ account for the effect of noise and stimulus cross-correlations. In the small rates and small correlations limit, we show here that we recover the main result of this paper. First we notice that in this limit:
\begin{align}
    \nu_{i j} &= \frac{r_{i j}^{\rm s}}{\sqrt{C_{i i}C_{j j}}} = \frac{r_{i j}^{\rm s}}{\sqrt{\mu_i\mu_j}}, \\
    \gamma_{i j}({\bf s}) &= \frac{\rho_{i j}^{\rm n}({\bf s})}{\sqrt{ C_{i i}({\bf s}) C_{j j}({\bf s})}} = \frac{\rho_{i j}^{\rm n}({\bf s})}{\sqrt{\mu_i({\bf s})\mu_j({\bf s})}}.
\end{align}
Replacing these expressions in $I^{t t, {\rm (1)}}_{i j}$, $I^{t t, {\rm (2)}}_{i j}$ and $I^{t t, {\rm (3)}}_{i j}$ then expanding the logarithms at small $r_{i j}^{\rm s}$ and $\rho_{i j}^{\rm n}({\bf s})$ (and thus $r_{i j}^{\rm n}$) and truncating at second order gives:
\begin{align}
\frac{t^2}{2} I^{t t, {\rm (1)}}_{i j} &= \mu_{i}\mu_{j}\left[\frac{r_{i j}^{\rm s}}{\sqrt{\mu_i\mu_j}}-\left(1+\frac{r_{i j}^{\rm s}}{\sqrt{\mu_i\mu_j}}\right)\ln \left({1+\frac{r_{i j}^{\rm s}}{\sqrt{\mu_i\mu_j}}}\right)\right] \nonumber \\
&\approx - \frac{1}{2}{r_{i j}^{\rm s}}^2 \label{eq:Itt1_2cd}, \\ \nonumber \\
\frac{t^2}{2} I^{t t, {\rm (2)}}_{i j} &= -\left\langle\rho_{i j}^{\rm n}({\bf s}) \sqrt{C_{i i}({\bf s})C_{j j}({\bf s})} \right\rangle_{s} \ln\left({1+\frac{r_{i j}^{\rm s}}{\sqrt{\mu_i\mu_j}}}\right) \nonumber \\
&= - C_{i j}^{\rm n} \ln\left({1+\frac{r_{i j}^{\rm s}}{\sqrt{\mu_i\mu_j}}}\right) \nonumber \\
&\approx - {r_{i j}^{\rm n}}{r_{i j}^{\rm s}} \label{eq:Itt2_2cd}, \\ \nonumber \\
\frac{t^2}{2} I^{t t, {\rm (3)}}_{i j} &= \bigg\langle\left(\mu_{i}(\mathbf{s}) \mu_{j}(\mathbf{s})+\rho_{i j}(\mathbf{s}) \sqrt{C_{i i}({\bf s})C_{j j}({\bf s})} \right) \times \bigg[ \ln{\left(1+\frac{\rho_{i j}^{\rm n}({\bf s})}{\sqrt{\mu_i({\bf s})\mu_j({\bf s})}}\right)} \nonumber \\ 
&~~~- \ln\left( 1 + \frac{\langle \rho_{i j}(\mathbf{s^{\prime}}) \sqrt{C_{i i}({\bf s^{\prime}})C_{j j}({\bf s^{\prime}})} \rangle_{\bf s^{\prime}}}{ \langle \mu_i({\bf s^{\prime}}) \mu_j({\bf s^{\prime}}) \rangle_{\bf s^{\prime}} } \right) \bigg] \bigg\rangle_{\bf s} \nonumber \\
& = \bigg\langle\left(\mu_{i}(\mathbf{s}) \mu_{j}(\mathbf{s})+ C_{i j}^{\rm n}({\bf s}) \right) \times \bigg[ \ln{\left(1+\frac{\rho_{i j}^{\rm n}({\bf s})}{\sqrt{\mu_i({\bf s})\mu_j({\bf s})}}\right)} - \ln\left( 1 + \frac{C_{i j}^{\rm n}}{C_{i j}^{\rm s} + \mu_i \mu_j} \right) \bigg] \bigg\rangle_{\bf s} \nonumber \\
& = \bigg\langle\left(\mu_{i}(\mathbf{s}) \mu_{j}(\mathbf{s})+ C_{i j}^{\rm n}({\bf s}) \right) \times \ln{\left(1+\frac{\rho_{i j}^{\rm n}({\bf s})}{\sqrt{\mu_i({\bf s})\mu_j({\bf s})}}\right)}\bigg\rangle_{\bf s} \nonumber \\
&~~~ - \mu_i \mu_j \left(1 + \frac{r_{i j}^{\rm s}}{\sqrt{\mu_i \mu_j}} + \frac{r_{i j}^{\rm n}}{\sqrt{\mu_i \mu_j}} \right) \times \ln\left( 1 + \frac{r_{i j}^{\rm n}/\sqrt{\mu_i \mu_j}}{r_{i j}^{\rm s}/\sqrt{\mu_i \mu_j} + 1} \right) \nonumber \\
& \approx \frac{1}{2}\left< {\rho_{i j}^{\rm n}({\bf s})}^2 \right>_{\bf s} - \frac{1}{2}{r_{i j}^{n}}^2. \label{eq:Itt3_2cd}
\end{align}
Summing up these contributions gives:
\begin{align}\label{Itt_expanded}
\frac{t^2}{2} I^{t t} &= \frac{t^2}{2} \sum_{i j} I^{t t, {\rm (1)}}_{i j} + I^{t t, {\rm (2)}}_{i j} + I^{t t, {\rm (3)}}_{i j} \nonumber \\
&= \frac{t^2}{2} \sum_{i} I^{t t, {\rm (1)}}_{ii} + t^2 \sum_{i<j} I^{t t, {\rm (1)}}_{i j} + I^{t t, {\rm (2)}}_{i j} + I^{t t, {\rm (3)}}_{i j} \nonumber \\
& \approx \frac{1}{2} \sum_{i}{\mu_{i}}^2\left[\nu_{i i}+\left(1+\nu_{i i}\right) \ln \left(\frac{1}{1+\nu_{i i}}\right)\right] - \frac{1}{2}\sum_{i<j} \left( (r_{i j}^{\rm s} + {r_{i j}^{n}})^2 - \left< {\rho_{i j}^{\rm n}({\bf s})}^2 \right>_{\bf s} \right) \nonumber \\
& \approx \frac{1}{2} \sum_{i}{\mu_{i}}^2\left[\nu_{i i}+\left(1+\nu_{i i}\right) \ln \left(\frac{1}{1+\nu_{i i}}\right)\right] - \frac{1}{2}\sum_{i<j} \left( {\rho_{i j}^{\rm tot}}^2 - \left< {\rho_{i j}^{\rm n}({\bf s})}^2 \right>_{\bf s} \right),
\end{align}
and finally:
\begin{equation}
\begin{split}\label{Itt_final}
I & \approx t I^{t} + \frac{t^2}{2} I^{t t} \\
& \approx I_{\rm single} + I_{\rm 2nd},
\end{split}
\end{equation}
with the single site term given by:
\begin{equation}
I_{\rm single} = \sum_{i}\left\langle\mu_{i}(\mathbf{s}) \ln \frac{\mu_{i}(\mathbf{s})}{\mu_{i}}\right\rangle_{s} + \frac{1}{2} \sum_{i}{\mu_{i}}^2\left[\nu_{i i}+\left(1+\nu_{i i}\right) \ln \left(\frac{1}{1+\nu_{i i}}\right)\right],
\end{equation}
and the second order cross-correlations contribution by:
\begin{equation}
I_{\rm 2cd} = - \frac{1}{2}\sum_{i<j} \left( {\rho_{i j}^{\rm tot}}^2 - \left< {\rho_{i j}^{\rm n}({\bf s})}^2 \right>_{\bf s} \right).
\end{equation}
In the small correlations and small rates (or Poisson) limit, the contribution to the mutual information of cross-correlations as described by the small time bin expansion coincides with the second order approximation derived in this paper.
Besides, this small correlation expansion allows us to interpret the different components of $I^{t t}$ directly in terms of pairwise correlations: First, we see from Eq. \ref{eq:Itt1_2cd} that $t^2 \sum_{i < j} I^{t t, {\rm (1)}}_{i j}$ contains the systematically detrimental effect of stimulus correlations in the response. Second, Eq. \ref{eq:Itt2_2cd} shows that $t^2 \sum_{i < j} I^{t t, {\rm (2)}}_{i j}$ quantifies the effect of interactions between stimulus and noise correlations, in accordance with the ``sign-rule" \cite{Hu14, Azeredo21}. Finally, Eq. \ref{eq:Itt3_2cd} shows how $t^2 \sum_{i < j} I^{t t, {\rm (3)}}_{i j}$ accounts for the beneficial effects of noise correlations beyond the ``sign-rule". In particular, we can observe that pairwise noise correlations need not fluctuate for $t^2 \sum_{i < j} I^{t t, {\rm (3)}}_{i j}$ to be positive (see App.~A\SI{} for a proof).

\end{document}